\documentclass[twocolumn,showpacs,preprintnumbers,amsmath,amssymb,aps]{revtex4}
\usepackage{graphicx}
\usepackage{dcolumn}
\usepackage{bm}

\def\ket#1{\mathinner{|{#1}\rangle}}

\begin{document}

\bibliographystyle{osajnl}

\title{Hyperfine spectroscopy using co-propagating pump-probe beams}
\author{Alok K. Singh, Sapam Ranjita Chanu, Dipankar Kaundilya and Vasant Natarajan}
\email{vasant@physics.iisc.ernet.in}
\homepage{www.physics.iisc.ernet.in/~vasant}
\affiliation{Department of Physics, Indian Institute of
Science, Bangalore 560\,012, INDIA}

\begin{abstract}
We have shown earlier that hyperfine spectroscopy in a
vapor cell using co-propagating pump-probe beams has many
advantages over the usual technique of saturated-absorption
spectroscopy using counter-propagating beams. The main
advantages are the absence of crossover resonances, the
appearance of the signal on a flat (Doppler-free)
background, and the higher signal-to-noise ratio of the
primary peaks. Interaction with non-zero-velocity atoms
causes additional peaks, but only one of them appears
within the primary spectrum. We first illustrate the
advantages of this technique for high-resolution
spectroscopy by studying the $D_2$ line of Rb. We then use
an acousto-optic modulator (AOM) for frequency calibration
to make precise hyperfine-interval measurements in the
first excited $P_{3/2}$ state of $^{85,87}$Rb and
$^{133}$Cs.
\end{abstract}

\pacs{32.10.Fn,42.62.Fi,42.50.Gy}


\maketitle

\section{Introduction}

High-resolution laser spectroscopy has been revolutionized
in the last two decades with the advent of low-cost tunable
diode lasers \cite{WIH91}. These diodes, when placed in an
external cavity with optical feedback, have frequency
uncertainty of about 1 MHz, which is small enough for
hyperfine transitions in atoms to be resolved
\cite{MSW92,BRW01}. Hyperfine spectroscopy, particularly in
the low-lying electronic states of alkali-metal atoms,
plays an important role in fine-tuning atomic wavefunctions
used in theoretical calculations. This is because
comparison between theoretical and experimental
determinations of hyperfine structure provides a stringent
test of atomic calculations in the vicinity of the nucleus
\cite{SJD99}. In addition, hyperfine structure in these
multielectron atoms is sensitive to core polarization and
core correlation effects \cite{AIV77}.

Many of the alkali-metal atoms have transitions to the
first-excited state (so-called $D$ lines) which are in the
near infrared, and therefore accessible with diode lasers.
They also have a high-enough vapor pressure near room
temperature that the spectroscopy can be done in a vapor
cell. The thermal motion of the atoms inside the cell
causes {\it Doppler broadening}, which is typically 100
times larger than the natural width of the hyperfine
transitions. The standard technique to overcome the
first-order Doppler effect is to use a {\it
counter-propagating} pump beam to saturate the transition
for zero-velocity atoms, in what is called
saturated-absorption spectroscopy (SAS) \cite{DEM82}.

Most atoms also have several closely-spaced hyperfine
levels within the Doppler profile. In these cases, it is
well known that SAS also produces spurious crossover
resonances in between each pair of hyperfine transitions.
They occur because, for some non-zero-velocity group, the
pump drives one transition while the probe drives the
other. In earlier work \cite{BAN03}, we have shown that the
use of co-propagating pump and probe beams overcomes the
problem of crossover resonances. Closely-spaced levels that
are not resolved in SAS can be resolved by this technique.
Probe transmission in such multilevel atoms is caused by
the phenomenon of electromagnetically induced transparency
(EIT) \cite{HAR97,DAN05}, and population depletion due to
optical pumping. Furthermore, by scanning only the pump
beam, the signal appears on a flat background without the
underlying Doppler profile seen in SAS. This is
advantageous for applications such as laser locking or
laser frequency measurement.

In this work, we present a complete study of the spectra
taken with the co-propagating technique in the $D_2$ line
of the two isotopes of Rb. In particular, we show that the
effect of non-zero velocity groups is to cause additional
peaks. However, in contrast to SAS, only one of these
spurious peaks appears within the spectrum, and the real
peaks are unaffected. Interestingly, the spurious peak
within the spectrum is almost negligible for transitions
starting from the upper ground hyperfine level, but highly
prominent for transitions starting from the lower level.
Such differences between the two levels have also been seen
in EIT, arising from the fact that the closed transition is
$F \rightarrow (F+1)$ for the upper hyperfine set and $F
\rightarrow (F-1)$ for the lower hyperfine set.

For a second set of experiments, we have used a single
laser along with an acousto-optic modulator (AOM) to
produce the pump-probe beams with a precisely-controlled
frequency offset. We then obtain the entire spectrum by
scanning the frequency of the AOM. The scan axis is
guaranteed to be linear because it is determined by the
frequency of the rf oscillator driving the AOM. A curve fit
to the observed spectrum yields the hyperfine interval. The
measurements have an accuracy of 20 kHz, which is
comparable to the accuracy of other techniques.

In recent work from our laboratory, we have reported
high-accuracy values for the hyperfine constants in the $D$
lines of all alkali atoms \cite{DAN08}. The hyperfine
intervals in that work were obtained (with an accuracy of 6
kHz) by {\it locking} the AOM to the neighboring
transition. One of the uncertainties when locking the AOM
is whether the lock point is exactly at the center of the
peak, since any shift would cause a systematic error in the
measurement. Though we had done experiments to verify that
this error was less than 2~kHz, we wanted to repeat the
measurements with another technique that was not at all
susceptible to errors arising from lock-point uncertainty.
As we will see below, the co-propagating technique achieves
precisely this. In addition, by measuring the entire
spectrum and looking at the symmetry of the line shape, we
can be sure that other sources of error are not
significant. The current set of measurements in $^{85}$Rb
and $^{133}$Cs, although having slightly smaller precision,
are consistent with our earlier work.

\section{Spectroscopy on the $D_2$ line of rubidium}
\begin{figure}
\centering{\resizebox{0.95\columnwidth}{!}{\includegraphics{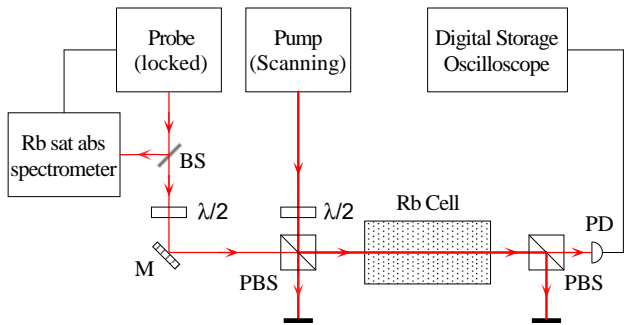}}}
\caption{Schematic of the experiment.  Figure key: BS --
beamplitter, $\lambda/2$ -- halfwave retardation plate, M
-- mirror, PBS -- polarizing beamsplitter cube, PD --
photodiode.}
 \label{schema}
\end{figure}

The schematic for the first set of experiments is shown in
Fig.\ \ref{schema}. The pump and probe beams are derived
from two home-built frequency-stabilized diode laser
systems \cite{BRW01} operating on the 780 nm $D_2$ line of
Rb. The linewidth of the lasers after stabilization is of
the order of 1 MHz. The output beams are elliptical with
$1/e^2$ size of $2 \times 4$ mm and powers of around 10
$\mu$W each. Part of the probe laser is sent into a Rb SAS
cell and the laser is locked to a hyperfine transition
using fm modulation spectroscopy. The pump laser is scanned
around the same set of transitions. The beams are mixed in
a polarizing beamsplitter cube (PBS) and {\it copropagate}
through a room-temperature vapor cell (5 cm long) with
orthogonal linear polarizations. Halfwave retardation
plates in the path of each beam allow precise control of
their powers. The probe beam is separated using a second
PBS, and its transmitted signal is detected with a
photodiode. The PBS's have extinction ratios of $1000:1$,
ensuring good purity of the detected signal.

\subsection{Spectrum in $^{87}$Rb, $F=2 \rightarrow F'=1,2,3$}
\begin{figure}
(a)\centering{\resizebox{0.9\columnwidth}{!}{\includegraphics{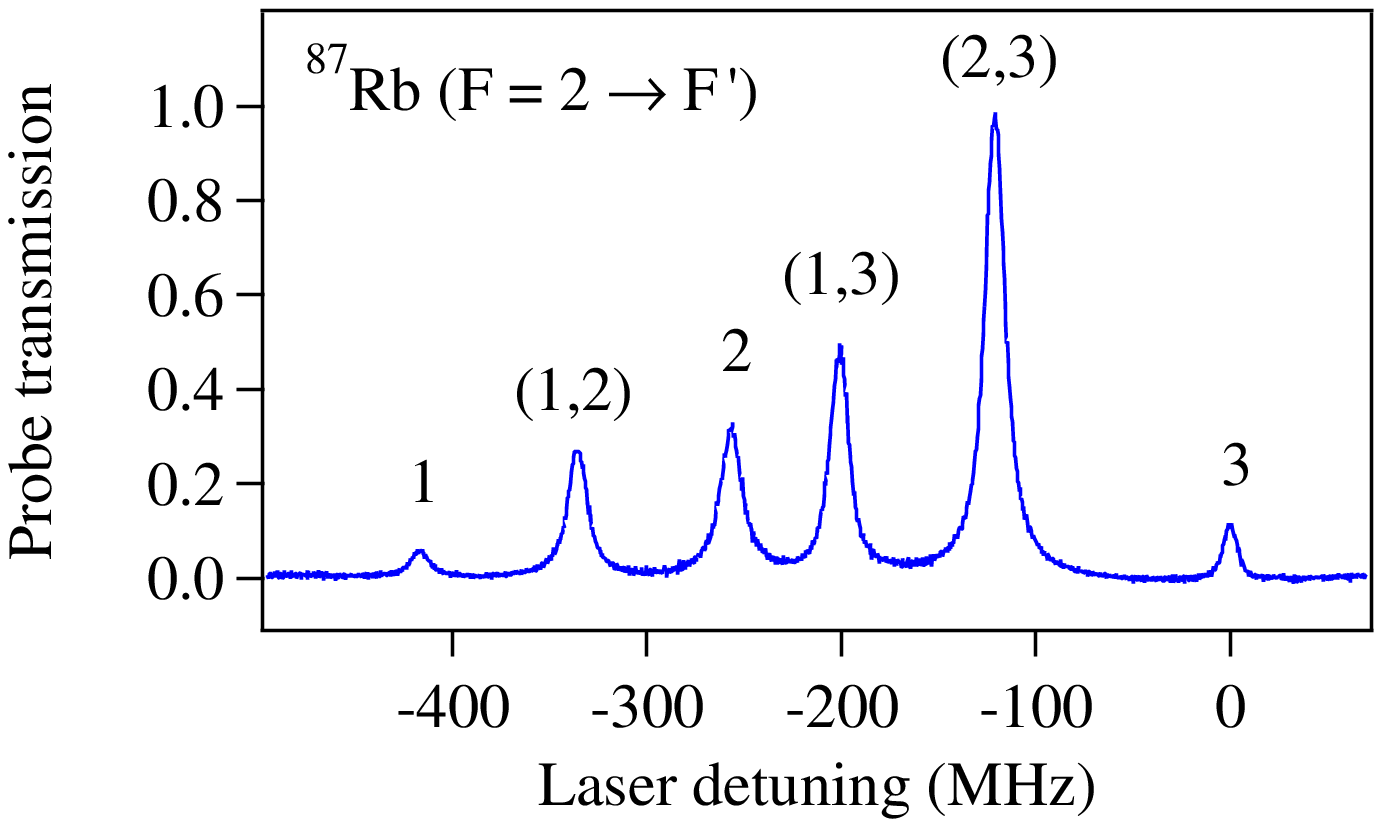}}} \\
(b)\centering{\resizebox{0.9\columnwidth}{!}{\includegraphics{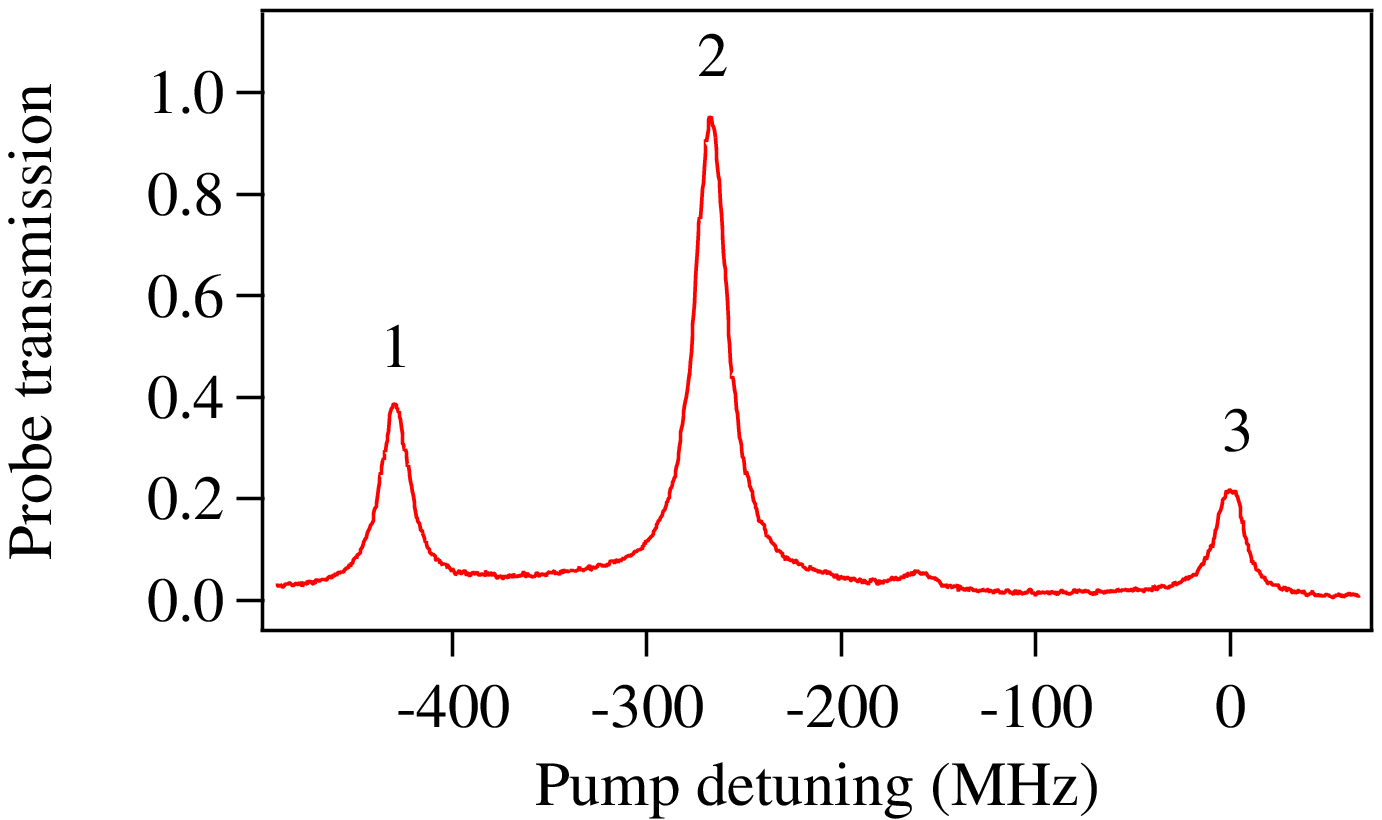}}} \\
(c)\centering{\resizebox{0.9\columnwidth}{!}{\includegraphics{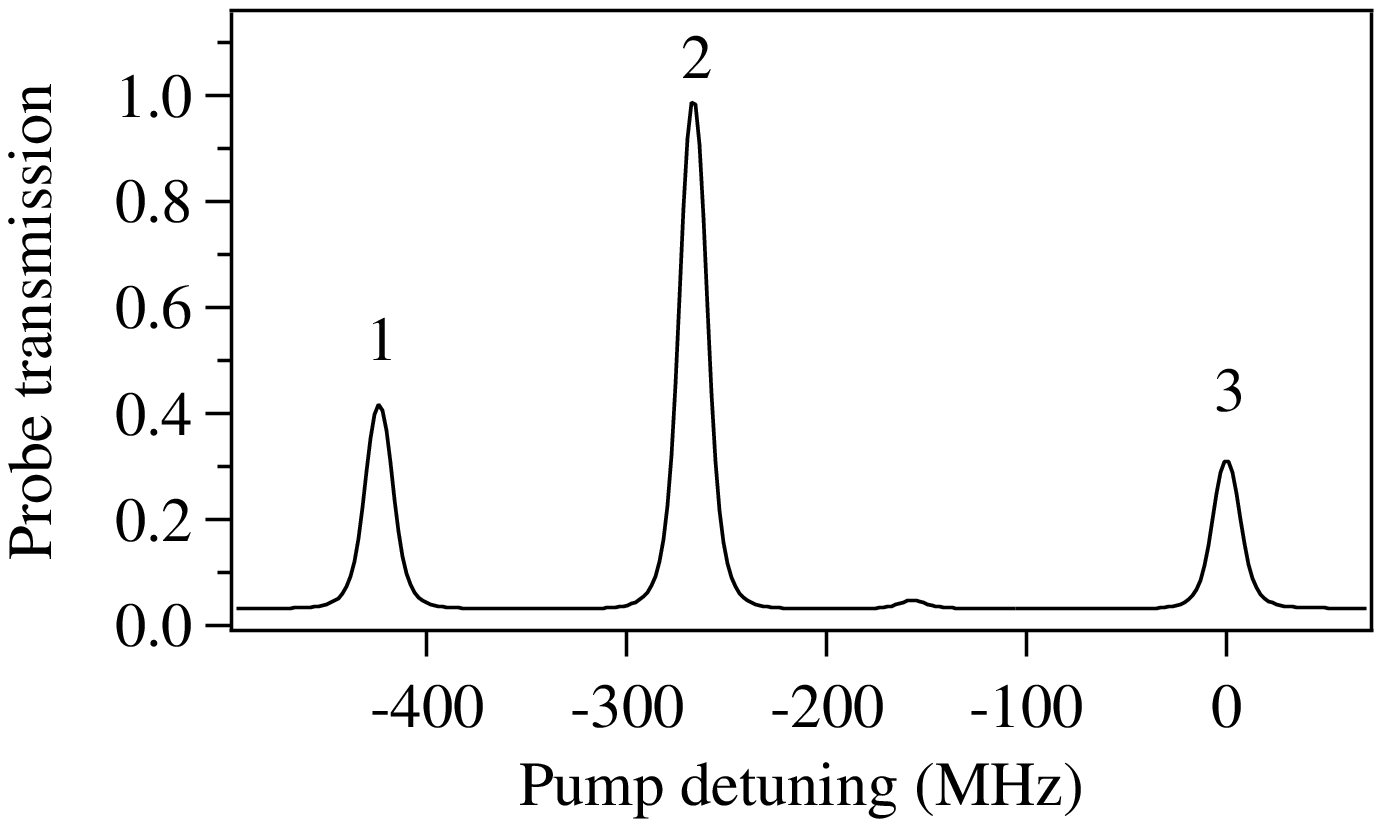}}}
\caption{(a) Doppler-subtracted saturated-absorption
spectrum for $F=2 \rightarrow F'$ transitions in $^{87}$Rb.
The hyperfine transitions are labelled with the value of
$F'$, and the crossover resonances in between with the two
values of $F'$. (b) Spectrum taken with co-propagating
pump-probe beams, with probe locked to the $F=2 \rightarrow
3$ transition and pump scanning. Note the absence of
crossover resonances and the better signal for the
hyperfine peaks. There is one additional peak at $-157$
MHz, as explained in the text. (c) Calculated spectrum
taking into account the full thermal velocity
distribution.}
 \label{satabs}
\end{figure}

As mentioned before, the main advantage of the
co-propagating configuration is the absence of spurious
crossover resonances. This difference is seen clearly in
Fig.\ \ref{satabs}. In (a), we show the usual
saturated-absorption spectrum for the $F=2 \rightarrow
F'=1,2,3$ transitions in $^{87}$Rb. The spectrum is Doppler
corrected, which is necessary because probe absorption
through a vapor cell will show a broad Doppler profile when
the probe addresses a velocity group different from that
resonant with the pump. When the pump and probe are
resonant with the same velocity class, we get transmission
peaks. As expected, there are three hyperfine peaks and
three crossovers. The crossovers are more prominent than
the actual peaks because two velocity classes contribute to
each crossover resonance, compared to one (zero-velocity)
class for each hyperfine peak. Probe transparency is
primarily caused by two effects: (i) saturation of
absorption caused by the strong pump beam, and (ii) optical
pumping into the $F=1$ ground hyperfine level for open
transitions (i.e.\ those involving the $F'=1$ and 2 excited
levels). In addition, there will be population
redistribution among the magnetic sublevels, which can
cause increased absorption or transparency depending on the
$F$ values of the levels. The linewidth of the peaks in the
figure is about 12 MHz, compared to the natural linewidth
of 6 MHz. This increase is typical in SAS and arises due to
a misalignment angle between the beams and power
broadening.

Now let us consider the spectrum shown in (b) taken with
the co-propagating configuration. The probe is locked to
the $F=2 \rightarrow 3$ transition and the pump is scanned
across the set of $F=2 \rightarrow F'=1,2,3$ transitions.
Since the probe is locked, its transmitted signal primarily
corresponds to absorption by zero-velocity atoms (i.e.\
atoms moving perpendicular to the laser beam) making
transitions to the $F'=3$ level. The signal remains flat
(or Doppler free) until the pump also comes into resonance
with a transition for the same zero-velocity atoms. Thus
there are three transmission peaks at the locations of the
hyperfine transitions, with no crossover resonances in
between. The hyperfine peaks are located at $-423.600$ MHz,
$-266.657$ MHz, and 0 \cite{DAN08}, all measured with
respect to the frequency of the locked probe laser. The
linewidth of the peaks is about 19 MHz, which is only 50\%
larger than the linewidth obtained in the
saturated-absorption spectrum. The primary cause for the
transparency peaks is the phenomenon of EIT in this V-type
system. The pump laser causes an {\it AC Stark shift of the
ground level} (creation of dressed states \cite{COR77}) and
hence reduces probe absorption at line center. In addition,
there are effects of saturation and optical pumping, but
these are less important than the EIT effects.

Since the experiments are done in a vapor cell with the
full Maxwell-Boltzmann distribution of velocities, we have
to consider that there will be two additional velocity
classes that absorb from the locked probe: both moving in
the same direction as the probe but with velocities such
that one drives transitions to the $F'=2$ level (266.657
MHz lower) and the second to the $F'=1$ level (423.600 MHz
lower). Each of these will cause three additional
transparency peaks from the mechanisms discussed above. The
first velocity class moves at 208 m/s and will cause peaks
at $-156.943$ MHz, 0, and $+266.657$ MHz, i.e., a set of
peaks shifted up by 266.657 MHz. The second velocity class
moves at 330 m/s and will cause peaks at 0, $156.943$ MHz,
and $+423.600$ MHz, i.e., a set of peaks shifted up by
423.600 MHz. Thus there will be 7 peaks in all, with 3 real
peaks and 4 spurious ones. However, only the peak at
$-156.943$ MHz will appear within the spectrum, caused by
the probe driving the $F=2 \rightarrow F'=2$ transition and
the pump driving the $F=2 \rightarrow F'=1$ transition. The
other three spurious peaks will lie outside the spectrum to
the right. This is indeed what is observed in Fig.\
\ref{satabs}(b): there is a small peak at $-157$ MHz within
the spectrum.

\begin{figure}
\centering{\resizebox{0.95\columnwidth}{!}{\includegraphics{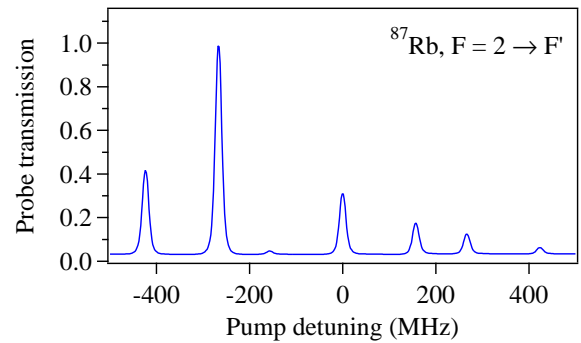}}}
\caption{Calculated spectrum extending on the other side of
the locked probe peak. There are now 7 peaks, as explained
in the text.}
 \label{theory}
\end{figure}

The above explanation is borne out by the calculated
spectrum shown in Fig.\ \ref{satabs}(c). Using a
density-matrix formulation, we can calculate the absorption
of a probe laser in a V-type system \cite{DAN05}. The
calculation is done for multiple hyperfine levels with full
thermal averaging. The only adjustable parameters are the
relative amplitudes of the three EIT resonances. As seen
from the figure, the calculation reproduces the locations
of the peaks in the measured spectrum. The observed
linewidth is slightly larger than the calculated one, but
this could be because of a small misalignment angle between
the beams, which is known to broaden the EIT resonance
\cite{CAT04}. The calculation shows that there is only one
spurious peak within the spectrum. However, if we extend
the calculation up to $+500$~MHz, we see all the seven
peaks mentioned in the previous paragraph. The extended
calculation is shown in Fig.\ \ref{theory}.

\begin{figure}
(a)\centering{\resizebox{0.9\columnwidth}{!}{\includegraphics{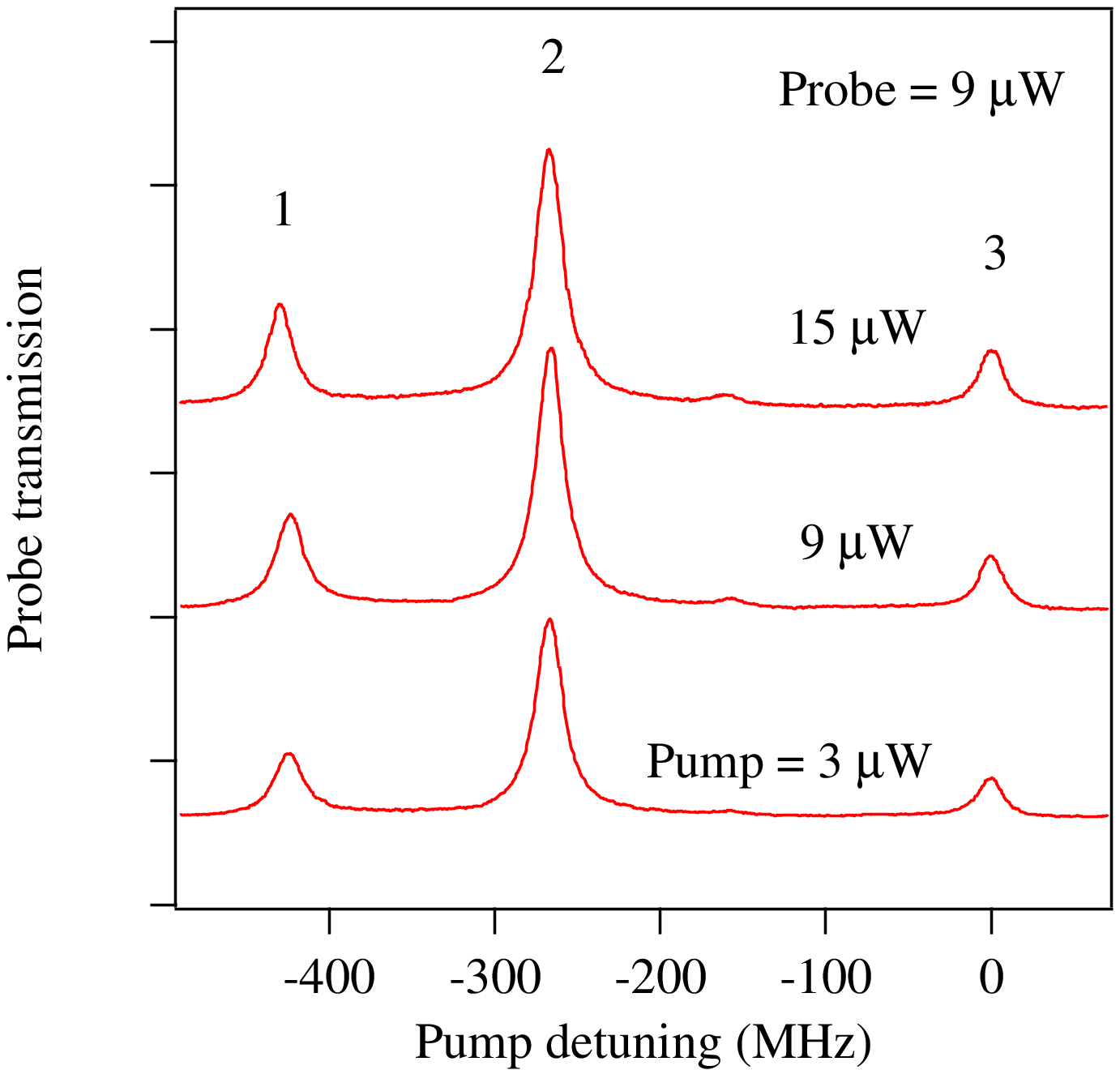}}} \\
(b)\centering{\resizebox{0.9\columnwidth}{!}{\includegraphics{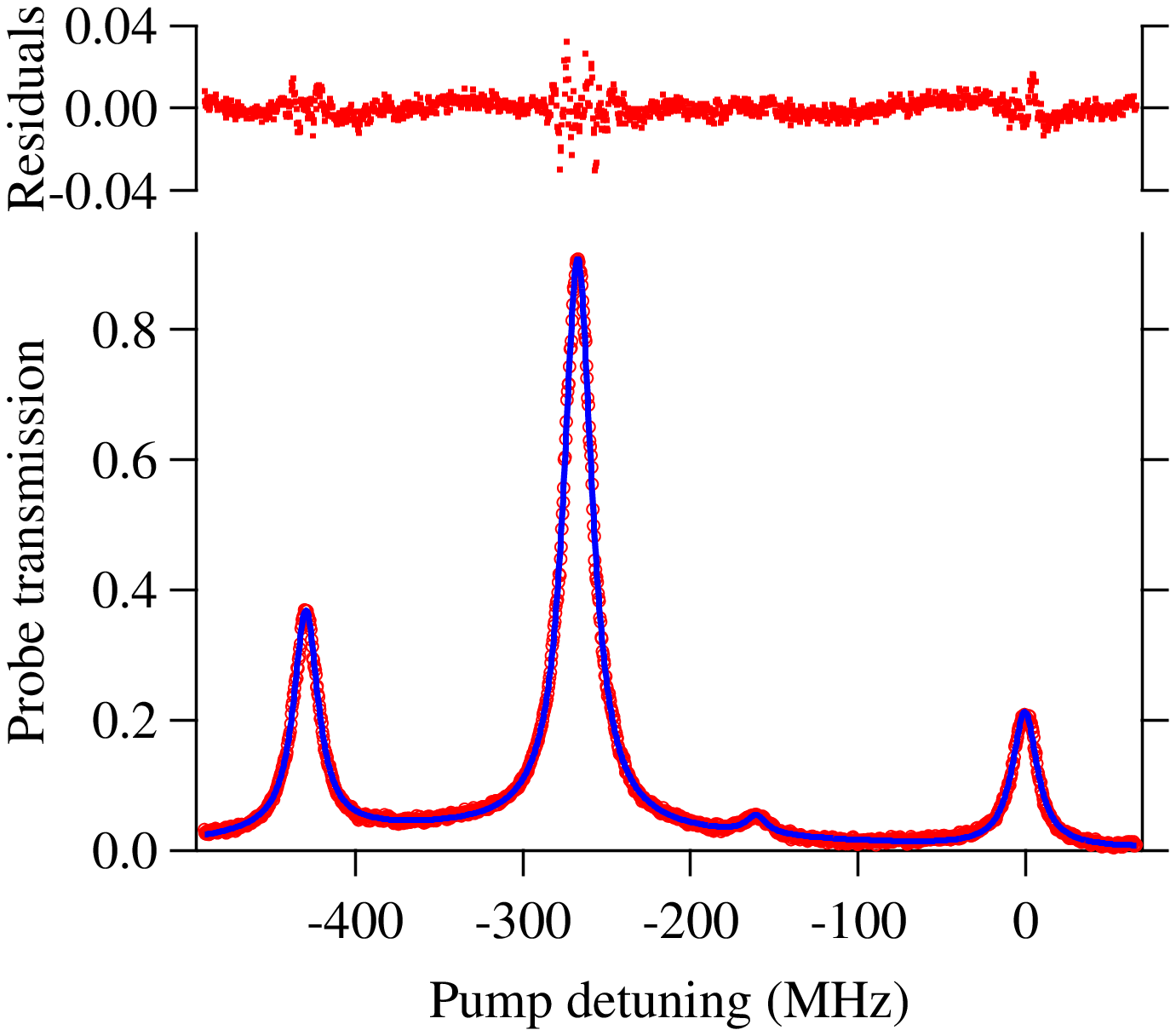}}}
\caption{(a) Spectra taken with co-propagating pump-probe
beams at three values of pump power. The probe beam is
locked to the $F=2 \rightarrow 3$ transition and the pump
beam is scanning. (b) Spectrum taken at the pump power of
15 $\mu$W shown with a multipeak Lorentzian fit. The fit
residuals are on top.}
 \label{coprop}
\end{figure}

The advantages of this scheme are quite clear from Fig.\
\ref{satabs}(b). The spectrum appears on a flat background,
obviating the need for Doppler subtraction as in the case
of SAS. There are no crossover resonances, which often
swamp the true peaks. And there is only one additional peak
within the spectrum due to absorption by non-zero velocity
atoms. In Fig.\ ref{coprop}, we show the effect of pump
power on the peaks. As the power is varied from $0.33$ to
$1.66$ times the probe power, the three main peaks remain
quite prominent with good signal-to-noise ratio. The
additional peak at $-157$ MHz increases in height, but not
significantly. By comparison, a good saturated-absorption
spectrum requires the pump-probe power ratio to be
accurately controlled to a value of $3:1$, with loss in
signal at lower pump powers and power broadening at higher
powers. In Fig.\ \ref{coprop}(b), we show a multipeak
Lorentzian fit to the spectrum measured with a pump power
of 15 $\mu$W. The residuals show that the line shape of all
the peaks is Lorentzian.

\subsection{Spectrum in $^{87}$Rb, $F=1 \rightarrow F'=0,1,2$}
\begin{figure}
\centering{\resizebox{0.95\columnwidth}{!}{\includegraphics{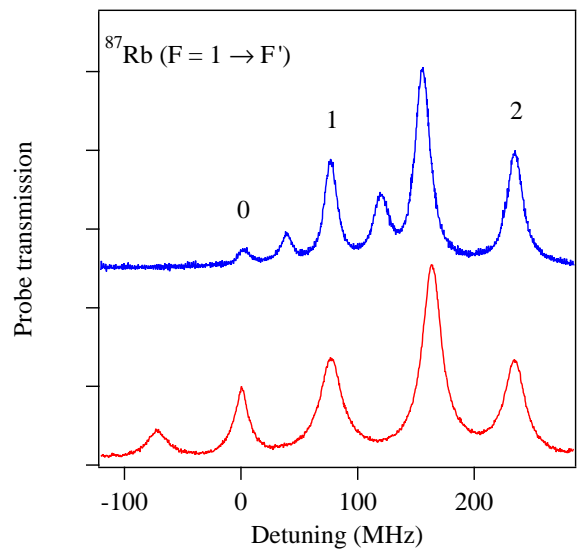}}}
\caption{Comparison of spectra with the two techniques for
lower-level $F=1 \rightarrow F'$ transitions in $^{87}$Rb.
The upper trace is the Doppler-subtracted
saturated-absorption spectrum, while the lower trace is
obtained with the probe locked to the $F=1 \rightarrow 0$
transition and pump scanning.}
 \label{rb10}
\end{figure}

The same advantages are seen in the spectrum of transitions
starting from the lower ground hyperfine level ($F=1
\rightarrow F'=0,1,2$) shown in Fig.\ \ref{rb10}. The
Doppler-subtracted saturated-absorption spectrum on top has
6 peaks including the 3 crossover resonances. The spectrum
with the co-propagating beams shown below is taken with the
probe locked to the $F=1 \rightarrow 0$ transition. The
beam powers are 9 $\mu$W (probe) and 15 $\mu$W (pump). It
appears on a flat background and shows the 3 hyperfine
peaks without any crossovers in between. The 3 hyperfine
peaks are located at 0, $+72.223$ MHz, and $+229.166$ MHz.
Two of the additional peaks are seen, one at $-72.223$ MHz
(outside the spectrum) and the other at $+156.943$ MHz
(within the spectrum). The additional peak within the
spectrum is due to atoms moving with a velocity of 52~m/s
such that the probe drives the $F=1 \rightarrow F'=1$
transition and the pump drives the $F=1 \rightarrow F'=2$
transition. This spurious peak is more prominent compared
to transitions starting from the upper ground level [see
Fig.\ \ref{satabs}(b)] (though the real peaks still have
high signal-to-noise ratio). The difference arises due to
the fact that the closed transition for this set is the
$F=1 \rightarrow F'=0$ transition, which has fewer magnetic
sublevels in the excited state compared to the ground
state. This leads to population trapping in the $m_F = \pm
1$ sublevels, and the relative importance of EIT effects in
causing probe transparency (which is the same for all
peaks) increases.

\subsection{Spectrum in $^{85}$Rb, $F=3 \rightarrow F'=2,3,4$}

\begin{figure}
\centering{\resizebox{0.9\columnwidth}{!}{\includegraphics{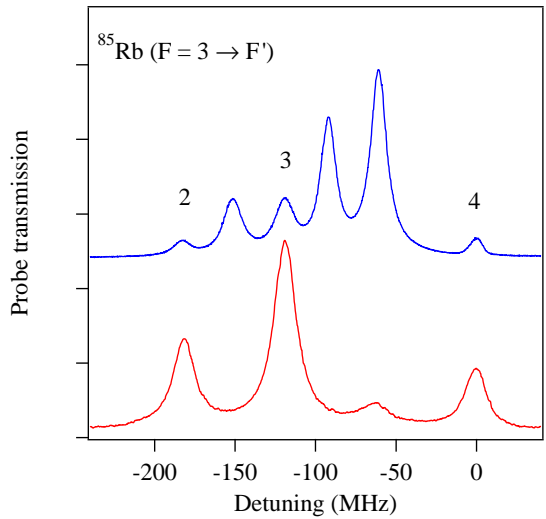}}}
\caption{Comparison of spectra with the two techniques for
upper-level $F=3 \rightarrow F'$ transitions in $^{85}$Rb.
The upper trace is the Doppler-subtracted
saturated-absorption spectrum, while the lower trace is
obtained with the probe locked to the $F=3 \rightarrow 4$
transition and pump scanning.}
 \label{rb34}
\end{figure}

The improvement with this technique is much more dramatic
in the spectra of the other isotope, $^{85}$Rb. For
transitions starting from the upper ground level ($F=3
\rightarrow F'=2,3,4$) shown in Fig.\ \ref{rb34}, the
hyperfine peaks corresponding to $F'=2$ and 4 in the
saturated-absorption spectrum are barely visible. In the
co-propagating spectrum shown below, the peaks become
prominent. The hyperfine intervals \cite{DAN08} are such
that the real peaks are at $-184.390$ MHz, $-120.966$ MHz,
and 0, while the additional peaks are at $-63.424$ MHz,
$+63.424$ MHz, $+120.966$ MHz, and $+184.390$ MHz. The
additional peak within the spectrum (at $-63.424$ MHz) is
almost negligible, as was observed for upper-level
transitions in $^{87}$Rb. The beam powers are 9 $\mu$W
(probe) and 15 $\mu$W (pump).

\subsection{Spectrum in $^{85}$Rb, $F=2 \rightarrow F'=1,2,3$}
\begin{figure}
\centering{\resizebox{0.9\columnwidth}{!}{\includegraphics{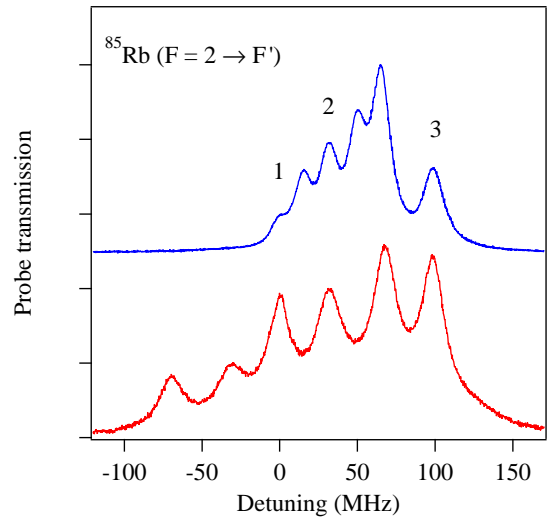}}}
\caption{Comparison of spectra with the two techniques for
lower-level $F=2 \rightarrow F'$ transitions in $^{85}$Rb.
The upper trace is the Doppler-subtracted
saturated-absorption spectrum, while the lower trace is
obtained with the probe locked to the $F=2 \rightarrow 1$
transition and pump scanning.}
 \label{rb21}
\end{figure}

For transitions starting from the lower ground level ($F=2
\rightarrow F'=1,2,3$) shown in Fig.\ \ref{rb21}, the
hyperfine interval between $F'=1$ and 2 is so small that
the crossover resonance in the saturated-absorption
spectrum completely swamps the $F'=1$ peak. However, the
spectrum with the co-propagating technique shows the peak
well resolved. The real peaks are located at 0, $+29.268$
MHz, and $+92.692$ MHz, while the additional peaks are at
$-92.692$ MHz, $-63.424$ MHz, $-29.268$ MHz, and $+63.424$
MHz \cite{DAN08}. The additional peak within the spectrum
(at $+63.424$ MHz) is quite prominent as in the case of
transitions starting from the lower hyperfine level in
$^{87}$Rb, again because the closed $F=2 \rightarrow F'=1$
transition has population trapping in the $m_F = \pm 2$
sublevels. There are two additional peaks appearing outside
the spectrum to the left, which are closer because of the
smaller hyperfine intervals. The beam powers are 9 $\mu$W
(probe) and 15 $\mu$W (pump).

\section{Hyperfine measurements using an acousto-optic modulator}
The above experiments were done using separate pump and
probe lasers. However, it is possible to do the experiment
with just one laser by using an AOM to produce the scanning
pump beam. The scan range of an AOM is limited to about 20
MHz, but this is large enough to scan across a hyperfine
peak. The main advantage of using an AOM is that the
frequency-scan axis (with respect to the probe beam) is
both linear and calibrated by the rf frequency of the
driver powering the AOM, thus allowing the hyperfine
interval to be measured accurately. By measuring the entire
peak, potential systematic errors due to locking of the
pump laser to a peak are avoided. In addition, if there is
a systematic shift in the lock point of the probe laser,
this will not cause an error in the interval because the
frequency which brings the pump into resonance will also be
shifted by the same amount, and hence the AOM offset (with
respect to the probe) for the spectrum will remain the
same.

We have therefore used a single laser and a scanning AOM to
measure hyperfine intervals in the $D_2$ lines of Rb and
Cs. The measurements are motivated by the fact that there
are several experimental values reported in the literature
that are somewhat discrepant from each other. In many
cases, we feel that a potential source of error is the
uncertainty in locking to a peak. In our current technique,
measuring the entire spectrum avoids such errors, as
discussed before.

\begin{figure}
\centering{\resizebox{0.95\columnwidth}{!}{\includegraphics{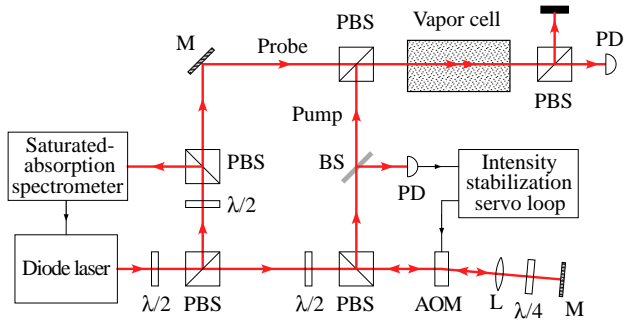}}}
\caption{Schematic of the experiment.  Figure key:
$\lambda/2$ -- halfwave retardation plate, PBS --
polarizing beamsplitter cube, AOM -- acousto-optic
modulator, L -- lens, $\lambda/4$ -- quarterwave
retardation plate, M -- mirror, BS -- beamplitter, PD --
photodiode.}
 \label{schema2}
\end{figure}

The experimental schematic for this second set of
experiments is shown in Fig.\ \ref{schema2}. As before, the
primary laser is a frequency-stabilized diode laser. The
probe beam is derived after locking the laser to a
hyperfine transition using SAS. The scanning pump beam is
frequency offset from the probe using a double-passed AOM.
The frequency is adjusted so that the pump is resonant with
a nearby hyperfine transition whose interval has to be
measured. The pump intensity is stabilized to better than
1\% in a servo-loop by controlling the rf power driving the
AOM. The two beams co-propagate through a vapor cell kept
inside a magnetic shield. The residual field (measured with
a three-axis fluxgate magnetometer) is below 5 mG. The
beams have orthogonal linear polarizations and are mixed
and separated using PBS's. The beam powers are about 15
$\mu$W each and adjusted to get good signal-to-noise ratio
in the spectrum.

The experiment proceeds as follows. An rf frequency
generator whose timebase is referenced to an ovenized
quartz clock (uncertainty less than $10^{-8}$) is used to
drive the AOM. A computer program is used to set the rf
frequency, and the probe signal is measured and recorded.
The frequency is changed in steps of 0.1 MHz over a range
of 15 MHz to obtain the complete spectrum. A curve fit to
the spectrum yields the AOM frequency at the peak center,
which is the hyperfine interval.

\subsection{Error analysis}
Systematic errors can arise due to one of the following
reasons.
\begin{enumerate}
\item[(i)] {\it Radiation-pressure effects.} Radiation
    pressure causes velocity redistribution of the
    atoms in the vapor cell. In the SAS technique, the
    opposite Doppler shifts for the counter-propagating
    beams can result in asymmetry of the observed
    lineshape. However, with co-propagating beams, the
    effects are less important because the Doppler
    shift will be the same for both beams and will not
    affect the hyperfine interval, similar to how the
    interval is insensitive to any detuning of the
    probe from resonance.

\item[(ii)] {\it Effect of stray magnetic fields.} The
    primary effect of a magnetic field is to split the
    Zeeman sublevels and broaden the line without
    affecting the line center. However, line shifts can
    occur if there is asymmetric optical pumping into
    Zeeman sublevels. For a transition $\ket{F,m_F}
    \rightarrow \ket{F',m_{F'}}$, the systematic shift
    of the line center is $\mu_B(g_{F'}m_{F'} -
    g_Fm_F)B$, where $\mu_B=1.4$ MHz/G is the Bohr
    magneton, $g$'s denote the Land\'e $g$ factors of
    the two levels, and $B$ is the magnetic field. The
    selection rule for dipole transitions is $\Delta m
    = 0,\pm 1$, depending on the direction of the
    magnetic field and the polarization of the light.
    Thus, if the beams are linearly polarized, there
    will be no asymmetric driving and the line center
    will not be shifted. We therefore minimize this
    error in two ways. First, we use polarizing cubes
    to ensure that the beams have near-perfect linear
    polarization. Second, we use a magnetic shield
    around the cell to minimize the field.

\end{enumerate}

The experiment is repeated by reversing the scan direction
to check for errors that might depend on which direction
the rf generator is scanned. Another source of error is
whether the intensity stabilization servo-loop stays
locked. But if this loses lock, it shows up in the spectrum
as an asymmetry of the line shape. Indeed, both the sources
of error discussed above also show up as asymmetry of the
line shape. Thus a symmetric line shape is a good
indication that the measurement proceeded correctly. From
the residual asymmetry, we estimate the systematic errors
to be 20~kHz.

\subsection{Measurements in the $5P_{3/2}$ state of $^{87}$Rb}
The first set of measurements were done in $^{87}$Rb. The
different values in the literature are consistent with each
other, and have an accuracy of 10 kHz. Therefore, our main
motivation was to see if the scanning-AOM technique worked
well and our error budget was proper.

\begin{figure}
\centering{\resizebox{0.9\columnwidth}{!}{\includegraphics{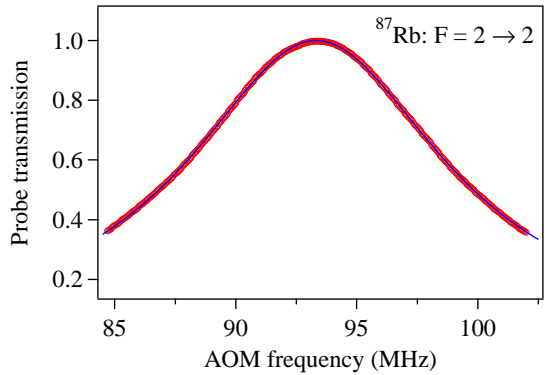}}}
\caption{Spectrum in $^{87}$Rb with probe locked to the
$F=2 \rightarrow 3$ transition, and pump scanning across
the $F=2 \rightarrow 2$ transition. The solid curve is a
fit to a Lorentzian lineshape.}
 \label{rb22}
\end{figure}

A typical spectrum with the probe locked to the $F=2
\rightarrow F'=3$ transition and pump scanning across the
$F=2 \rightarrow F'=2$ transition is shown in Fig.\
\ref{rb22}. We saw earlier that the line shape was well
described by a Lorentzian. We therefore fit a Lorentzian
curve to the spectrum and extract the peak center. Note the
symmetry of the spectrum and the high signal-to-noise
ratio. The $\{ 3-2 \}$ interval is twice the center
frequency (because the AOM is double passed). For technical
reasons, there is an additional AOM with a fixed frequency
in the path of the probe, and this offset has to be added
to obtain the interval.

The average value from 14 individual measurements is listed
in Table \ref{t1}. The standard deviation of the set is 32
kHz, which means the expected error in the mean is
$32/\sqrt{14} = 8.6$ kHz, less than our estimated error of
20 kHz. This value is compared to other values reported in
the literature. The two most accurate measurements
\cite{DAN08,YSJ96} have uncertainties below 10 kHz and have
overlapping error bars. The more recent measurement
\cite{DAN08} is also from our laboratory and used an AOM to
measure the interval, but the AOM was locked to the peak.
The current measurement obtained by measuring the entire
spectrum is consistent with this value, thus giving
confidence in the current technique. The only slightly
discrepant measurement is from the work in Ref.\
\cite{BGR91}, which is $1.7 \sigma$ away.

\begin{table}
\caption{Comparison of measurements of hyperfine intervals
in the $5P_{3/2}$ state of $^{87}$Rb to previous results.
The last row is calculated from the $A$ and $B$
coefficients reported therein. All values in MHz.}
\begin{ruledtabular}
\begin{tabular}{lc}
$\{ 3-2 \} $ Interval & Reference \\
\hline
266.653(20) & This work \\
266.657(8) & \cite{DAN08} \\
266.650(9) & \cite{YSJ96} \\
266.503(84) & \cite{BGR91} \\
\end{tabular}
\end{ruledtabular}
 \label{t1}
\end{table}

\subsection{Measurements in the $5P_{3/2}$ state of $^{85}$Rb}
\begin{figure}
(a)\centering{\resizebox{0.75\columnwidth}{!}{\includegraphics{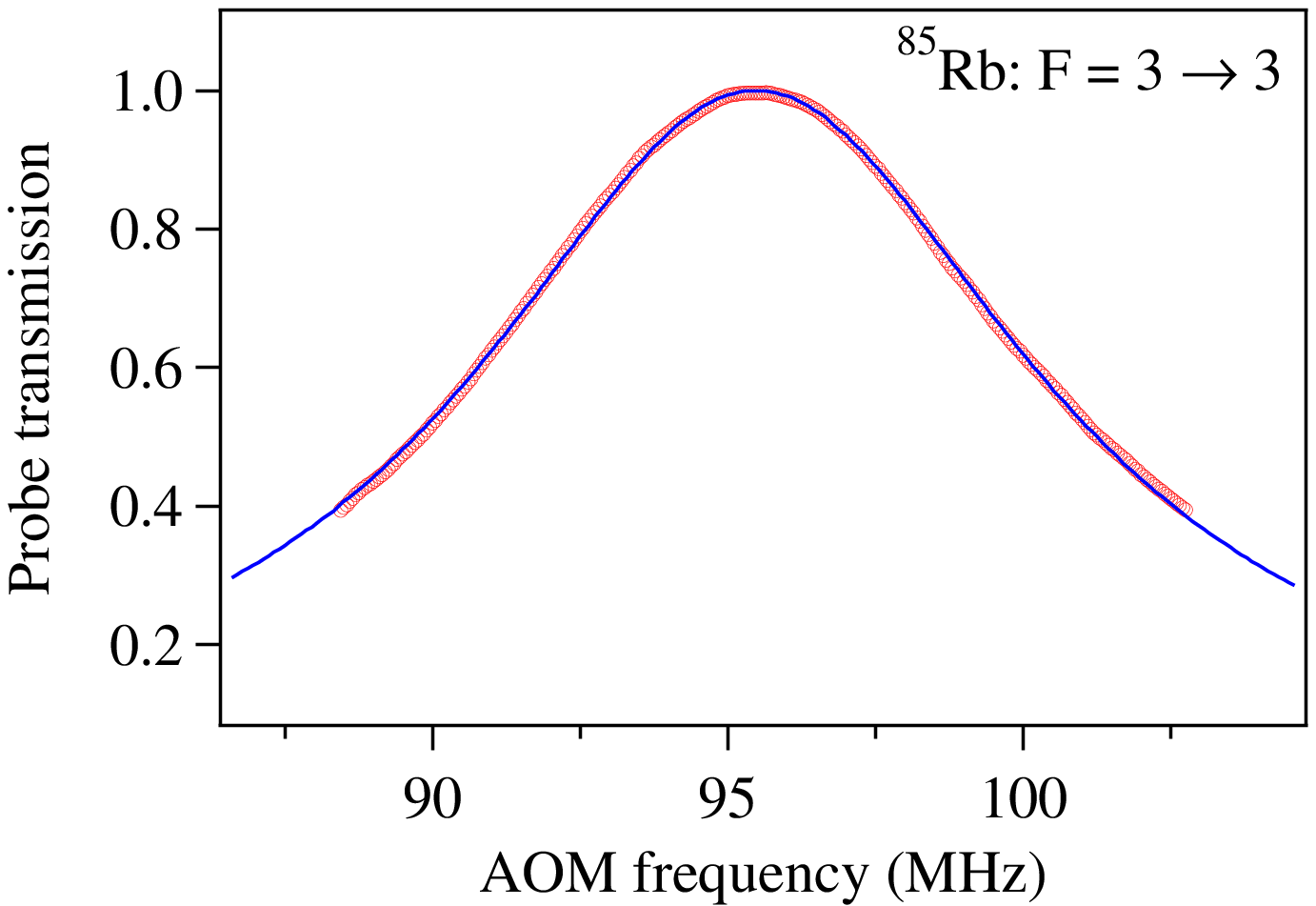}}}
(b)\centering{\resizebox{0.75\columnwidth}{!}{\includegraphics{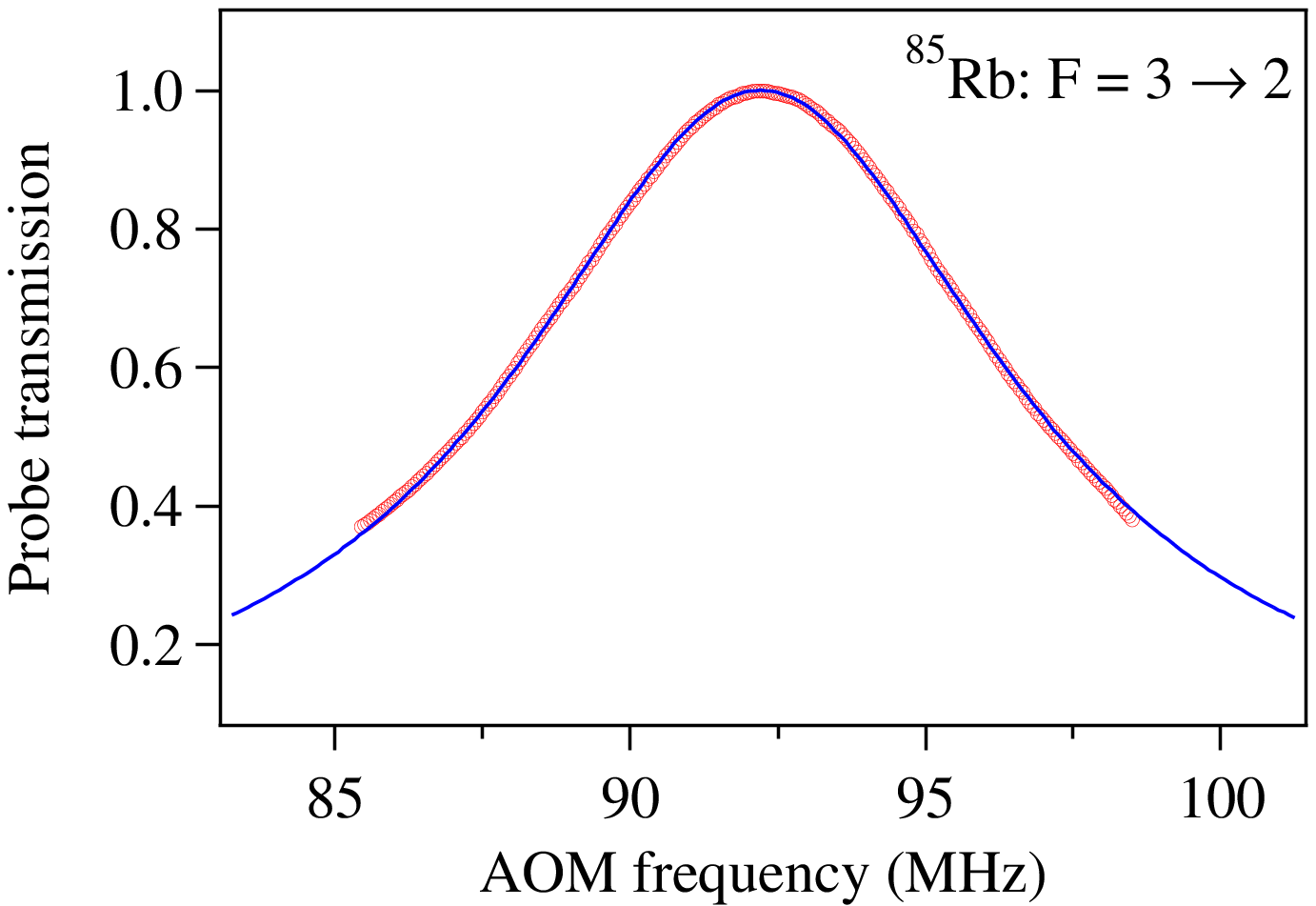}}}
\caption{Spectra in $^{85}$Rb with probe beam locked to the
$F=3 \rightarrow 4$ transition, and pump beam scanning
across (a) the $F=3 \rightarrow 3$ transition, and (b) the
$F=3 \rightarrow 2$ transition. The solid curves are
Lorentzian fits.}
 \label{rb33}
\end{figure}

With the reliability of the technique established with
measurements in $^{87}$Rb, we turned to the other isotope,
namely $^{85}$Rb. The probe was locked to the $F=3
\rightarrow F'=4$ transition, and the pump was scanned
either across the $F=3 \rightarrow F'=3$ transition or the
$F=3 \rightarrow F'=2$ transition. Typical spectra for the
two cases are shown in Fig.\ \ref{rb33}. As before,
Lorentzian fits to the measured spectra were used to
determine the peak center, and thus the $\{ 4-3 \}$ and $\{
3-2 \}$ intervals.

The average values for the two intervals are listed in
Table \ref{t2}. For the $\{ 4-3 \}$ interval, the standard
deviation from a set of 14 measurements is 35 kHz. For the
$\{ 3-2 \}$ interval, the standard deviation from 11
measurements is 34 kHz. These values are also compared to
other values in the literature. There are two
non-overlapping sets for the $\{ 4-3 \}$ interval. The
value from Ref.\ \cite{BGR91} is $3.7 \sigma$ away from the
other two values, which are both from our laboratory and
both of which relied on AOM locking. Ref.\ \cite{BGR91} is
also the work in which the value in $^{87}$Rb was
discrepant by $1.7 \sigma$. The current measurement is
consistent with our previous ones. All the values for the
$\{ 3-2 \}$ interval are consistent with each other.

\begin{table}
\caption{Comparison of measurements of hyperfine intervals
in the $5P_{3/2}$ state of $^{85}$Rb to previous results.
The last row is calculated from the $A$ and $B$
coefficients reported therein. All values in MHz.}
\begin{ruledtabular}
\begin{tabular}{llc}
$\{ 4-3 \} $ Interval & $\{ 3-2 \} $ Interval & Reference \\
\hline
120.958(20) & 63.436(20) & This work \\
120.966(8) & 63.424(6) & \cite{DAN08} \\
120.960(20) & 63.420(31) & \cite{RKN03} \\
120.506(124) &  63.402(93) & \cite{BGR91} \\
\end{tabular}
\end{ruledtabular}
 \label{t2}
\end{table}

\subsection{Measurements in the $6P_{3/2}$ state of $^{133}$Cs}
\begin{figure}
(a)\centering{\resizebox{0.75\columnwidth}{!}{\includegraphics{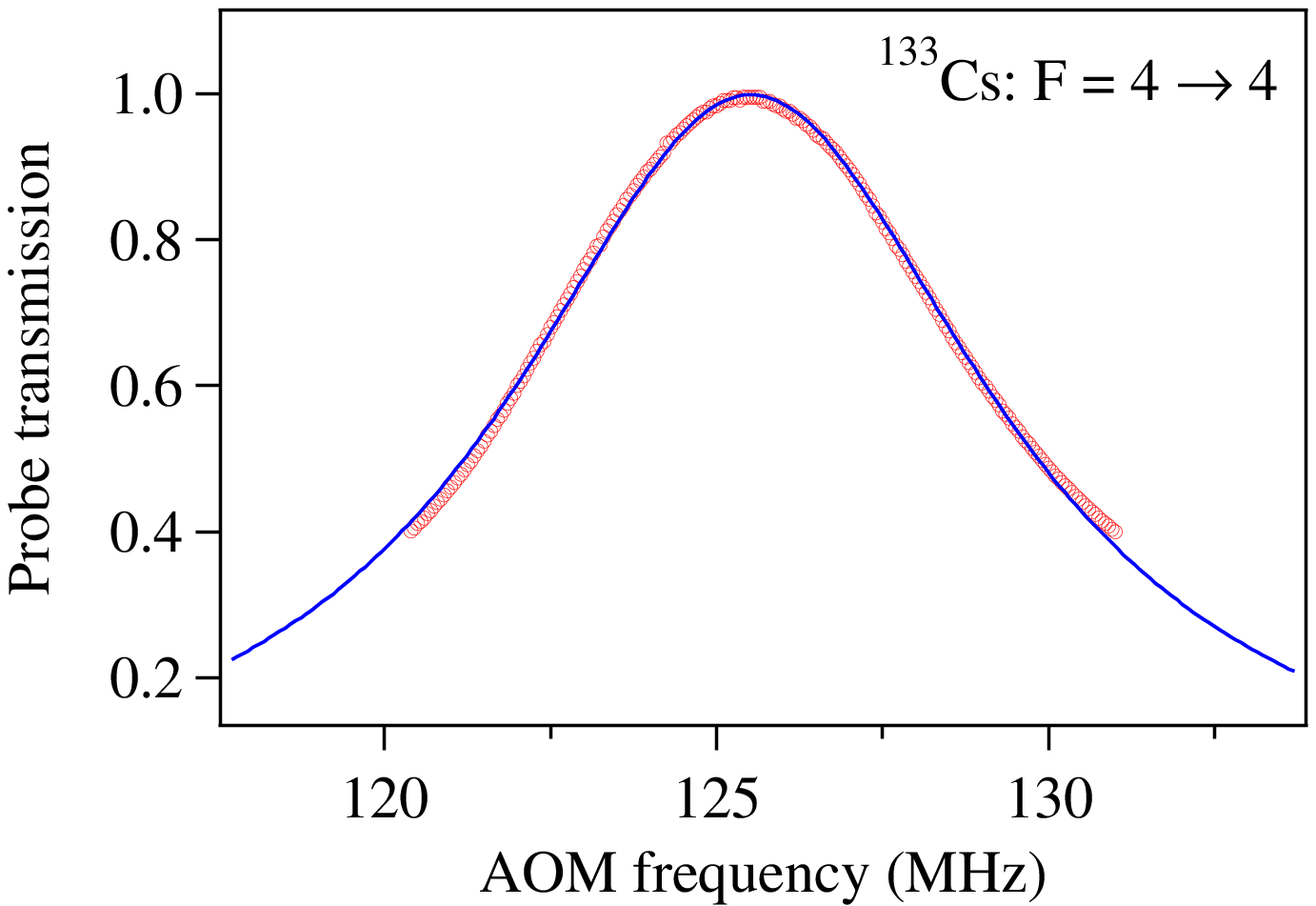}}}
(b)\centering{\resizebox{0.75\columnwidth}{!}{\includegraphics{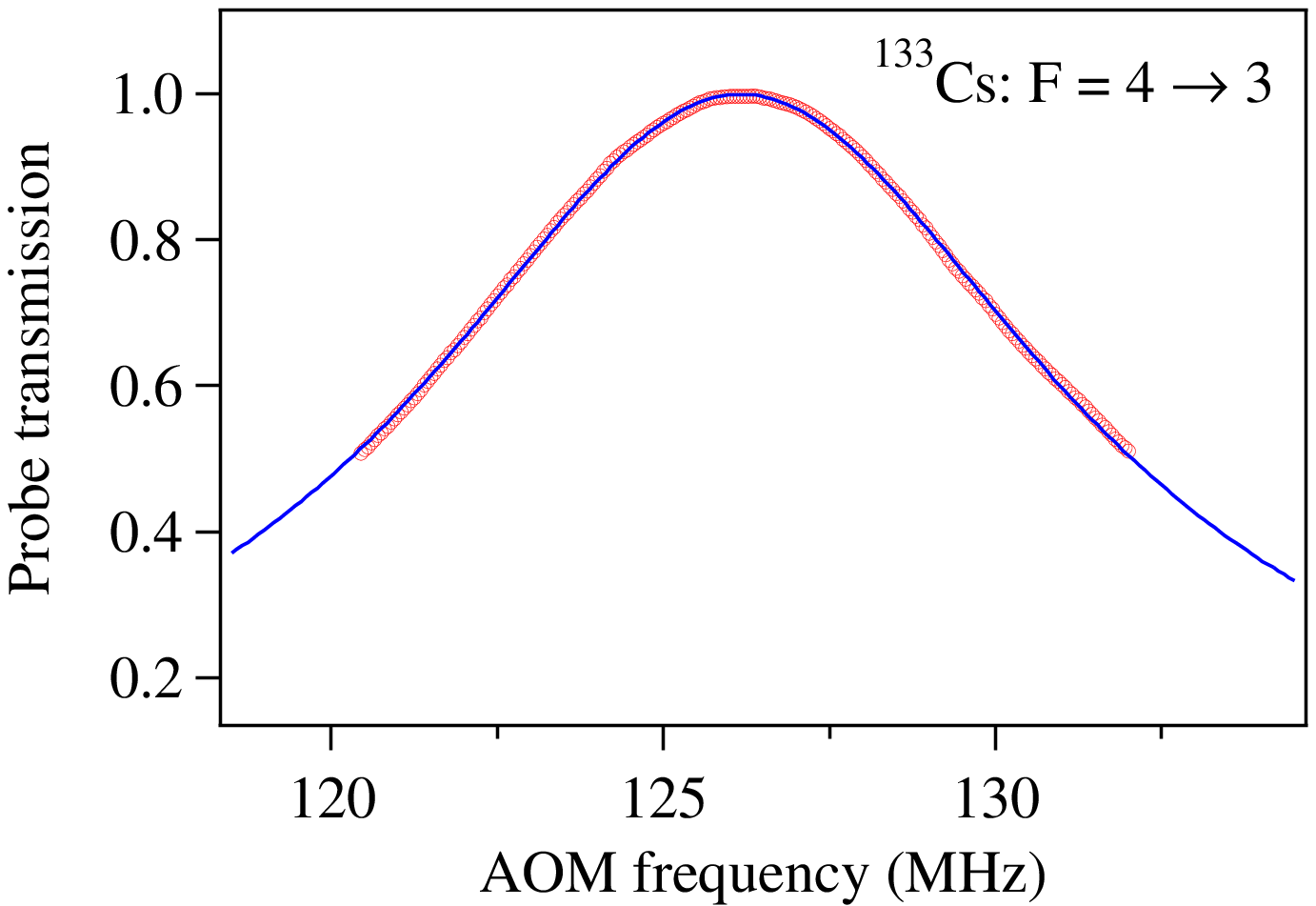}}}
\caption{Spectra in $^{133}$Cs with probe beam locked to
the $F=4 \rightarrow 5$ transition, and pump beam scanning
across (a) the $F=4 \rightarrow 4$ transition, and (b) the
$F=4 \rightarrow 3$ transition. The solid curve are
Lorentzian fits.}
 \label{cs44}
\end{figure}

The next set of measurements was done on the $D_2$ line in
$^{133}$Cs at 852 nm. For this, the probe was locked to the
$F=4 \rightarrow F'=5$ transition and the pump was scanned
either across the $F=4 \rightarrow F'=4$ transition or the
$F=4 \rightarrow F'=3$ transition. Representative spectra
for the two cases are shown in Fig.\ \ref{cs44}. Lorentzian
fits to the spectra yielded the line center and hence the
$\{ 5-4 \}$ and $\{ 4-3 \}$ intervals.

The average value for the $\{ 5-4 \}$ interval is
251.031(20) MHz, as listed in Table \ref{t3}. This was
obtained from a set of 27 independent measurements with a
standard deviation of 21 kHz. We concentrated on this
interval because two of the values reported in the
literature, 251.092(2) MHz from Ref.\ \cite{GDT03} and
251.000(20) from Ref.\ \cite{TAW88}, differ by $4.5
\sigma$. The more recent measurement of the two
\cite{GDT03} was done using a frequency comb. An earlier
measurement from our laboratory using the AOM locking
technique \cite{DAN05} yielded a result of 251.037(6) MHz,
which was consistent with the earlier value at the $1.5
\sigma$ level, but totally inconsistent with the
frequency-comb result (difference of $9\sigma$). Our
current value vindicates our earlier result since it is
consistent with the work in Ref.\ \cite{TAW88} but not with
the frequency-comb result.
\begin{table}
\caption{Comparison of measurements of hyperfine intervals
in the $6P_{3/2}$ state of $^{133}$Cs to previous results.
All values in MHz.}
\begin{tabular}{llc}
$\{ 5-4 \} $ Interval & $\{ 4-3 \} $ Interval & Reference \\
\hline
251.031(20) & 201.260(20) & This work \\
251.037(6) & 201.266(6) & \cite{DAN05} \\
251.092(2) & 201.287(1) & \cite{GDT03} \\
251.000(20) & 201.240(20) & \cite{TAW88} \\
\end{tabular}
 \label{t3}
\end{table}

For the $\{ 4-3 \}$ interval, we obtain an average value of
201.260(20) MHz from a set of 10 measurements with a
standard deviation of 33 kHz. The value from the work in
Ref.\ \cite{TAW88} was 201.240(20) MHz, while the more
recent frequency-comb work in Ref.\ \cite{GDT03} reported a
value of 201.287(1) MHz. The inconsistency of $2.4 \sigma$
is smaller but still quite significant. Again, our previous
result of 201.266(6) MHz obtained with AOM locking
\cite{DAN05} overlapped with the earlier value but was
inconsistent with the frequency-comb result. Our new value,
though with a larger error bar, gives confidence in the
previous measurement.

\section{Conclusions}
In summary, we have shown that hyperfine spectroscopy with
co-propagating beams in a vapor cell has several advantages
over conventional saturated-absorption spectroscopy. In
addition to the usual mechanisms responsible for probe
transparency, there are EIT effects that enhance the peaks,
which is supported by density-matrix calculations. As a
result, the primary peaks are more prominent and appear
with good signal-to-noise ratio. The transmitted signal
appears on a flat background (Doppler-free) and does not
have the problem of crossover resonances in between
hyperfine transitions (which are stronger and often swamp
the true peaks). Absorption by non-zero velocity groups
causes additional peaks, but only one of them appears
within the spectrum. These observations are again supported
by density-matrix calculations taking the thermal velocity
distribution into account. An important difference between
transitions starting from the upper ground level and
transitions starting from the lower ground level, is that
the additional peak is almost negligible in the first case
and quite prominent in the second case. This difference
arises because of the difference in number of magnetic
sublevels for the closed transition in each set.

We have adapted this technique to make measurements of
hyperfine intervals by using one laser along with an AOM to
produce the scanning pump beam. We measure intervals in the
$D_2$ lines of Rb and Cs with 20 kHz precision. By
measuring the entire spectrum and looking at the symmetry
of the line shape, we avoid several potential sources of
systematic error. The measurements are consistent with
earlier results from our laboratory obtained by locking the
AOM to the frequency difference, and show that our earlier
error budget was reasonable.

\begin{acknowledgments}
This work was supported by the Department of Science and
Technology, India. V.N. acknowledges support from the Homi
Bhabha Fellowship Council and A.K.S. from the Council of
Scientific and Industrial Research, India.
\end{acknowledgments}


\end{document}